\documentclass[]{spie}           
\usepackage[]{graphicx}

\title{A feasibility study of future observations with MIDI and other VLTI 
science instruments:\\
The example of the Galactic Center}

\author{Joerg-Uwe~Pott\supit{a,b}, Andreas~Glindemann\supit{a}, 
Andreas~Eckart\supit{b}, Markus~Sch{\"o}ller\supit{c}, 
Christoph~Leinert\supit{d}, Thomas~Viehmann\supit{b}, Massimo Robberto\supit{e}
\skiplinehalf
\supit{a}European Southern Observatory, Karl-Schwarzschild-Str.\ 2, D-85748 
Garching bei M\"unchen, Germany; \\
\supit{b}1st Institute of Physics, University 
of Cologne, Z\"ulpicher Str. 77, D-50937 K\"oln, Germany\\
Garching bei M\"unchen, Germany; \\
\supit{b}1st Institute of Physics, University 
of Cologne, Z\"ulpicher Str.\ 77, D-50937 K\"oln, Germany\\
\supit{c}European Southern Observatory, Casilla 19001, Santiago, 
Chile\\
\supit{d}Max-Planck-Institut f{\"u}r Astronomie, K{\"o}nigstuhl 17, D-69117 Heidelberg, Germany \\
\supit{e}STScI,3700 San Martin Drive, Johns Hopkins University Homewood Campus Baltimore, MD 21218 
USA
}

\authorinfo{Further author information: Send correspondence to J.-U.~Pott:\\ 
E-mail: jpott@eso.org, Telephone: +49-3200-6516}

\begin{document} 
  \maketitle 

\begin{abstract}
Interferometry with the Very Large Telescope Interferometer (VLTI) will allow imaging of the Galactic Center and the 
nuclei of extragalactic sources at an angular resolution of a few 
milliarcseconds. VLTI will be a prime instrument to study the 
immediate environment of the massive black hole at the center of the Milky Way. 
With the MID infrared Interferometric instrument (MIDI) for example the enigmatic compact dust embedded MIR-excess sources 
within the central parsec should be resolvable. Further the observations of external 
galactic nuclei will allow unprecedented measurements of physical parameters 
(i.e. structure and luminosity) in these systems. 
With the exception of a few 'self-referencing' sources these faint-target observations will benefit 
from the available off-axis wavefront-correction 
system STRAP, working on suitable guide stars (GS).\\
To fully exploit the use of VLTI within this context, the following questions have to be 
addressed among others: How feasible is blind-pointing on (faint) science targets? Are VLTI observations still efficiently feasible if these faint science targets exceed the usual angular distance ($\le$1 arcmin) to a GS candidate, enabling a standard 
closed-loop tip-tilt correction? How is the fringe-tracking procedure affected 
in densely populated regions such as the Galactic Center? What preparatory steps have to be performed to successfully observe these non-standard targets with the VLTI? \\
In this contribution, we present aspects for the preparation of VLTI observations, which will be conducted in the near future. Considering these example observations of the Galactic Center region, several details of observing modes are discussed, which are necessary to observe such science targets. The final goal is the 
definition of observational strategies that are optimized for the discussed 
classes of targets, which provide properties touching the limits of VLTI observability.\end{abstract}



\section{INTRODUCTION}
\label{sect:intro}  
The construction of the VLTI has progressed step by step over the past years. Since the recent start of the regular science observations with MIDI, it has entered the final phase: The usage of the VLTI as an ordinary telescope facility, providing the astronomer with the extraordinary capabilities of an optical/infrared interferometer. Although the first scientific results, based on the publicly available data of commissioning and science demonstration runs (for a summary see \cite{2003Msngr.114...26R}) are very encouraging, most of the observed targets are bright, isolated and compact sources, thus ideal targets for interferometry. 

There is still only few workaday experience made with fainter targets. E.g.\ a number of scientifically interesting targets (deeply embedded young or evolved stars, active galactic nuclei) are  too faint in the optical wavelength regime to enable the use of the Adaptive Optics devices STRAP and MACAO. In these cases the off-axis observation of a Guide Star is foreseen to enable wave-front correction and efficient target acquisition and tracking. But such a Guide Star is not always available. Also crowded fields with several sources within the field-of-view of the interferometric science instruments may bear difficulties for the standard observability and the applied fringe tracking mechanisms. Of especial interest for the general astronomical community is the question of how easy  and straight forward VLTI observations close to the limiting technical capabilities can be conducted.

Thus a feasibility study is needed to test the usage of the VLTI under such adverse conditions. As described in the following sections, the region of the Galactic Center provides numerous scientifically interesting targets, which fulfill one or more of these conditions. It is therefore reasonable to extract from foreseen scientific observations of the GC region with the VLTI information and experiences for such a feasibility study. The different science cases of these observations are presented, dependent on the foreseen VLTI beam combining instrument (MIDI or AMBER for the near-infrared). As a first result of this study the capabilities of VLTI/MIDI are discussed in detail with respect to the successful preparation of similar observations.


\section{VLTI on the Galactic Center} 
The Galactic Center is due to its proximity ($8\pm0.5~kpc$; 
\cite{1993ARA&A..31..345R} a unique region to observe the surroundings of a supermassive black hole 
(\cite{1996Natur.383..415E,2000Natur.407..349G,2002Natur.419..694S}). \cite{2002SPIE.4835...12E} 
have given a general overview of the scientific potential for interferometric observations of the 
center of our galaxy. Several issues can be addressed by 
the reduction of interferometric datasets at (near)infrared wavelengths. 
Even the most 
recently observed infrared flares from the accretion disk around the black hole itself (\cite{2003Natur.425..934G}) are within reach of the upcoming interferometers (the VLTI\footnote{{\bf V}ery {\bf 
L}arge {\bf T}elescope {\bf I}nterferometer at ESO's Paranal Observatory; {\em 
http://www.eso.org/projects/vlti/}}, the LBT\footnote{{\bf L}arge {\bf B}inocular {\bf T}elescope in 
Arizona, {\em http://medusa.as.arizona.edu/lbto/}} and the Keck Interferometer\footnote{atop Mauna 
Kea; {\em http://planetquest.jpl.nasa.gov/Keck/keck\_index.html}}).

A central question is the understanding of the (apparently ongoing) starformation in the 
central parsec. One scenario for the formation of the found massive stars is the infall of a massive 
dense cloud less than $10^7~yr$ ago (\cite{2001ApJ...546L..39G}). Such a cloud would have been highly compressed 
and became gravitationally unstable. \cite{1996Natur.382..602S} discussed within their article 
different morphologies of the molecular gas such as a clumpy disk and a spiral-arm like geometry. 
They showed that both can lead to the ongoing star formation in the Galactic Center as observed 
(e.g.\ \cite{1995ApJ...447L..95K}, \cite{1995ApJ...441..603B}).

\subsection{MIDI}
\begin{table}
\begin{center}
\begin{tabular}{l|l}
Wavelength coverage & N-band (8-13$\mu$m)\\
\hline
Limiting magnitude$^1$  &	1~Jy \\
Available baselines & UT2-UT3(47~m), UT3-UT4(63~m), UT2-UT4(89~m)	\\
Interferometric resolution in N-band & 40~mas with 50~m projected baseline length \\
Target acquisition FOV & 2~arcsec \\
Fringe spectral dispersion  & Prism (R=30),Grism (R=230) \\
Slit for spectral dispersion &	Width=0.52 arcsec on sky \\
Spectral filtering & entire N-band \\
Fringe acquisition mode	& Fourier Mode: 5 points/fringe, 10 fringes/scan (indicative)\\
Fringe tracking	 & Internal (coherencing from data in Fourier mode)\\
tip-/tilt correction & STRAP unit in Coud{\'e} focus of the UT's \\
tip-/tilt requirements & V$\le16$; off-axis (max 1') Guide Star  possible\\
\end{tabular}
\end{center}
\caption{\label{tab_MIDIvalues}The offered MIDI observing modes in ESO's Period~74 (Oct.'04-Mar.'05). More and most recent information on these continuously changing data are presented on the web: {\em http://www.eso.org/instruments/midi/index.html}.\newline
  $^1$: refers to the correlated flux of the observed target, i.e. the radiated flux density reduced by the interferometric visibility, which depends on the observing wavelength, baseline and the target's brightness distribution}
\end{table}
MIDI is the mid-infrared two-beam combining facility at the VLTI. It was developed by an international consortium under the leadership of the Max-Planck Institut f\"ur Astronomie in Heidelberg, Germany. The first commissioning phase has been finished successfully. The data of the Science Verification Programme are publicly available in the ESO data archive\footnote{{\em http://archive.eso.org/}}. And although the number of commissioned capabilities and observing modes are still limited with respect to the final variety of possible fields of application, already the first extragalactic target (NGC~1068) has been observed several times and led to intriguing results (\cite{2004Natur.429..47J}). 

In April 2004 for the first time the usage of MIDI was offered to the whole astrophysical community. To give an idea of the (at the moment) possible astrophysical applications, some key values are listed in Table~\ref{tab_MIDIvalues}. In the next section MIDI observations of sources in the Galactic Center are presented. The observations will be conducted in July 2004, when the GC is well visible from Cerro Paranal. Because these targets are close to MIDI's (current) instrumental requirements, valuable experience in observing faint targets, within a crowded field of several sources, and with possible position uncertainties will be made (cf. Sect.~\ref{sect_MIDIcaps}).

\subsubsection{Mid-infrared Sources} 

Beside the population of young massive HeI stars (cf. Sect.~\ref{sect_massivestars}) 
there have been found at least half a dozen dust embedded sources with featureless K-band spectra 
within the central parsec. The nature of these strong 10~$\mu$m sources (IRS 21, IRS 1W, IRS 10W, 
IRS 3, etc.) is rather unclear. Initially \cite{1985ApJ...299.1007G} suggested IRS~21 to be an 
externally heated, high-density dust clump. Until recently the observational findings let them 
appear to be young stellar objects. More properties were published, such as the strong polarization 
(17\% at 2~$\mu$m; \cite{1995ApJ...445L..23E,1995ApJ...447L..95K}), the MIR excess and featureless K-band 
spectra. Thus \cite{1995ApJ...447L..95K} and \cite{2001A&A...376..124C} proposed to classify IRS 21 rather 
as protostar or embedded early-type star. 

Recently published diffraction-limited $2-25~\mu$m images 
suggest that IRS~21 is possibly a Wolf-Rayet star (\cite{2002ApJ...575..860T}). They and 
\cite{Tanner2003} observed with single-telescope observations extended infrared structures, which not 
necessarily have to be explained by young stars, still embedded in their natal molecular clouds. To 
understand more about these enigmatic objects, it turned out, that it is crucial to investigate the 
connection between the intrinsic source and its surroundigs. Therefore mid-infrared observations at 
highest angular resolution are needed. 

During summer 2004 we will observe some of these targets with 
the mid-infrared interferometric beam-combiner MIDI of the VLTI. The results of this observing run 
will show, if a bow-shock like structure (\cite{Tanner2003} can be observed and confirmed in the 
N-band in a direct (interferometric) measurement, without deconvolution of single aperture images. 
From the orientation and size of this bowshock the relative motion of the embedded source within the 
(moving) surrounding gas/dust of the so called Northern Arm can be derived. 

The involved dynamical 
calculations are described in some detail in (\cite{2004ApJ...602..770G} and references within). They 
obtained with the GEMINI North Telescope high resolution images of IRS~8 which has 
comparable properties to the other infrared sources, embedded in the Northern Arm, but is located 
farther away ($\sim30"$ from GC) from the central cluster. They could derive from the morphology of 
the bowshock the stellar proper velocity. This can be done with similar calculations for the targets 
of our observations. 

Thus the precise observation of a bow-shock, produced by strong stellar winds of fast moving 
stars in a dense gaseous medium can complement or even replace (at lower accuracy) the dynamical 
information of emission lines, which are not present for the discussed sample. The study of objects 
like IRS~1W and IRS~3 helps us to understand the physical conditions at the 
Galactic Center. These are dominated by the presence of the massive black hole, the star formation, 
the evolution of massive stars, and the properties of the interstellar gas in the extreme 
environment of the Galactic center. Providing the bow-shock hypothesis, these sources compress the 
more tenuous gas and dust of the central streamers and may therefore have vice versa an influence on 
the star formation. 

\subsubsection{\label{sect_MIDIcaps}VLTI/MIDI capabilities, required and tested by the Galactic Center observations}

Several properties of IRS~1W, IRS~3 and the other dust embedded sources in the GC region are classifying the MIDI observations as non-standard ones, which therefore have to be carried out in visitor mode. Because of the lack of larger experience with such non-standard MIDI observations we present here as the example of our Galactic Center observations some topics of general interest for the observer of non-standard MIDI targets.

\begin{itemize}
\item The targets are only bright in the mid-infrared range. What does this imply for the observability?
\end{itemize}

All optical observations of the Galactic Center suffer from very high extinction at optical wavelengths (about 30~magnitudes) due to the high amount of interstellar dust along the line of sight. But also other interesting mid-infrared sources are often deeply dust-embedded and not visible in the optical, because mostly thermally heated dust and gas is radiating at 10~$\mu$m. This complicates both the target acquisition and the tip-/tilt correction, done by the STRAP unit in the Coud{\'e}-foci of the Unit Telescopes (see \cite{1998SPIE.3353..224B} for an introduction into the Adaptive Optics system of the VLTI). 

The best solution is given by a natural Guide Star within the 2~arcmin FOV of the UT's Coud{\'e}-focus\footnote{i.e. the limiting distance between science target and source is about 1~arcmin}, fulfilling the V$\le$16~mag requirement of the STRAP unit. Then the tip-/tilt correction ensures, that MIDI is illuminated constantly with a coordinate precision of $\sim 0.05~$arsec, whereas without correction the reimaged image will move around due to atmospheric turbulences and lower the coordinate precision down to about 0.2~arcsec. 

Fortunately there is such a Guide Star within reach of the Galactic Center. But several tests have shown, that also if the STRAP unit cannot be used to stabilize the image, the target can be successfully acquired by MIDI, if the target coordinates are precise enough. 

\begin{itemize}
\item How are deficient coordinates affecting the target acquisition?
\end{itemize}

This again may be an important issue for several MIDI observations, because often the mid-infrared data are relying on low-resolution observations\footnote{e.g.\ IRAS Sky Survey Atlas shows images with a resolution of 4-5~arcmin; cf. {\em http://irsa.ipac.caltech.edu/IRASdocs/issa.exp.sup/}}. Only most recently high-resolution mid-infrared imagers at 8-meter class telescopes are upcoming in the Southern Hemisphere (VISIR at ESO's VLT and T-Recs at GEMINI South). But both are still under comissioning or inaugurated most recently. 

Also without matching the STRAP requirements the VLT's field stabilization is tracking the source coordinates with $\le 0.1''$ accuracy. Therefore the major problem is the small FOV of 2~arcsec of the VLTI. If the deficiency of target coordinates are surpassing this limit, the target cannot be seen on the MIDI detector. Then a successful target acquisition can only be performed by a time consuming search on the sky, which is strongly exacerbated, if the science target is visible only at mid-infrared wavelenghts\footnote{the technical CCDs in the Nasmyth/Coud{\'e} foci of the UT's are blind in N-band}. This case should {\em always} be excluded by the astronomer by pre-imaging the source at the highest available resolution to avoid waste of the given time, which is especially valuable, because until now an uncalibrated visibility measurement takes 20-30~minutes.

Another stumbling block to a successful fringe measurement is the limiting magnitude requirement of the target itself (Tab.~\ref{tab_MIDIvalues}). 

\begin{itemize}
\item Does the {\em correlated} flux density of the target surpass 1~Jy?
\end{itemize}

\begin{figure}
\begin{center}
\includegraphics{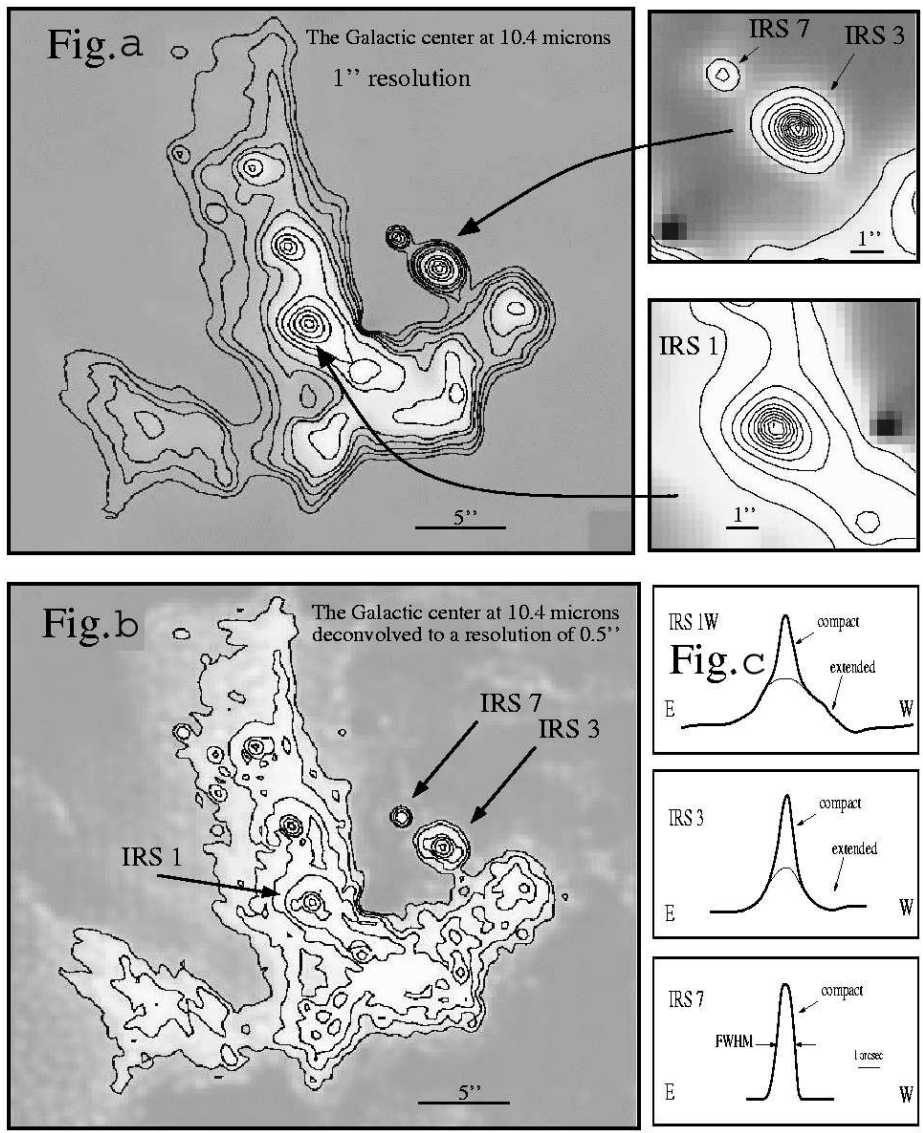}
\end{center}
\caption{\label{fig_MIDIprep}{\bf a:} 10.4$\mu$m  1~arcsec resolution image of the
Galactic Center obtained at the UKIRT (2.5, 3.75, 5, 7.5, 10, 17.5,
25, 37.5, 50, 75, 100 \% of the peak brightness). The total mid-infrared flux of IRS~1W, which contains within its extension of $\sim$2.5'' the brightest compact MIR source within the entire central cluster,  rises over the N-band from 35~Jy at 8.3~$\mu$m up to 71~Jy at 12.4$\mu$m (\cite{1985ApJ...299.1007G}). Panels to the right show the region around IRS1W and IRS3;
{\bf b:} The same image deconvolved to a resolution of about 0.5'' with the Richardson-Lucy deconvolution algorithm;
{\bf c:} RA cuts through IRS1W, IRS3, and IRS7 in the seeing limited image.
All three images show that there are compact $\le$0.3'' source components
at the centers of IRS1W and IRS3. Here we have indicated the compact components
in comparison to the unresolved source IRS~7. For IRS~1W and IRS~3 we show the
compact and extended flux distributions separately.}
\end{figure}

In general this issue leads to the same advice as the previous one: Only from modern high-resolution single-telescope observations there is a reasonable chance to derive the expected visibility at the needed level of accuracy to predict the correlated flux\footnote{which is the product of the visibility and the flux}. To explain this important observation preparation we present in Fig.~\ref{fig_MIDIprep} the procedure in some detail at the example of Galactic Center sources IRS~1W and IRS~3. The panel {\em c} shows linear cuts through the seeing limited images. The comparison with the totally unresolved source IRS~7 shows that our targets are composed of an extended and a compact component with respect to the UKIRT PSF at 10.4~$\mu$m. Components of arcsec-extension will be fully resolved by MIDI at all available baselines and will not contribute to the measured fringe contrast. The question of the correlated flux is transformed to the extension of the compact component. Of course single-telescope observations can only lead to upper limits of source size, detectable by MIDI. 

In our example both IRS~1W and IRS~3 contain compact components of less than 0.3'' angular size. 
Sometimes further hints on the MIR size of sources may be deducible from NIR data sets. But this has to be done very carefully and cannot replace a mid-infrared pre-imaging. In most cases the intrinsic physical radiation processes which account for the observed radiation are rather different in both wavelength-regimes.

The last step is to calculate the visibility from the expected source size. The online-tool {\em VLTI Visibility Calculator} is created for this task and is availble on the webpages of the European Southern Observatory\footnote{{\em http://www.eso.org/observing/etc/}}. 

\begin{figure}
\begin{center}
\includegraphics{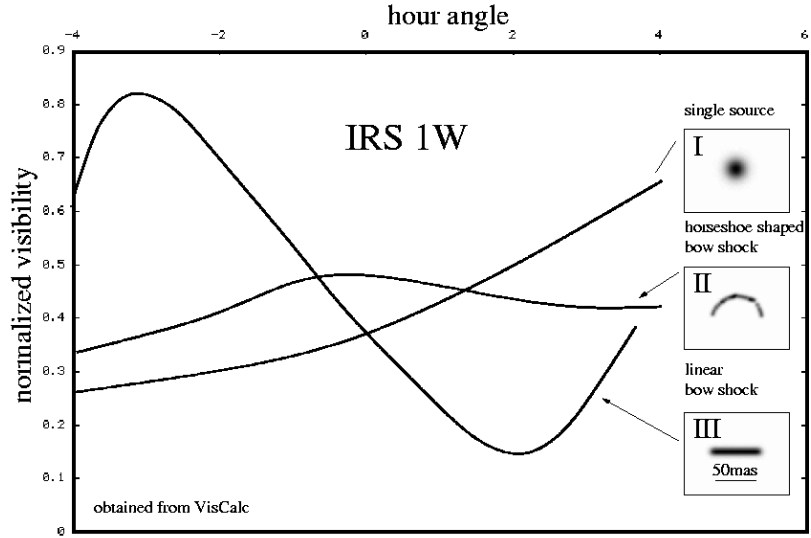}
\end{center}
\caption{\label{fig_MIDIvis}Calculation of the change of modelled visibilities with the hour angle. The used parameters are: Baseline UT2-UT3, source coordinates of IRS~1W and observation in July.}
\end{figure}
This is what the astronomer can do before the observation. If the brightness distribution of the source will result in fringe detections by MIDI, can only definitively be shown by the observation itself. To demonstrate to the general astronomer, presumably a novice in interferometry, how strongly the visibility (and therefore the interferometric detectability) still depends on the source structure, we calculated different scenarios of IRS~1W on the estimated source size order of magnitude (Fig.~\ref{fig_MIDIvis}). Further the figure shows immediately that {\em no} fringe detection at the beginning of the night not necessarily means: {\em no} fringes at all!
This fact always has to be taken into account to fully exploit the given MIDI time.

Further questions may be interesting and can be answered by the presented observations:
\begin{itemize}

\item How can the system handle crowded fields like the 
GC region, where several sources within the Coud{\'e}-field-of-view are present. 

\item Can the observer efficiently handle the complex VLTI system to rapidly switch between the scientific target and 
reference sources?  

\end{itemize}
At the current state of preparatory work, these questions cannot be answered extensively. But it was recently verified, that during the target-acquisition phase with closed tip-/tilt loop an R.A./Decl. offset within the 2~arcmin FOV of the Coud{\'e} foci can be performed easily within less than 10~min. And the observer can offset even further if the decreased image quality is accepted due to the loss of the tip-/tilt correction.

Finally two issues are interesting for the observation of faint MIR targets:

\begin{itemize}

\item How feasible/fruitful is a fringe measurement, if the tip/tilt correction in the Coud{\'e}-focus 
is not working? Either because of a lacking Guide Star or because of thin cirrus inhibiting the applicability of such a guide star.

\item And last but not least, how to use efficiently the MIDI time, if the correlated 
source flux is below the detection limit. 
\end {itemize}

\begin{table}
\begin{center}
\begin{tabular}{l|r|r|l}
Name	& Central wavelength ($\mu$m)&	FWHM ($\mu$m) &	Comment \\
\hline
$[ArIII]$	& 9.00	& 0.14	& PAH1 \\
$[SIV]$	& 10.48	& 0.16	& \\
N11.3	& 11.28	& 0.61	& PAH2 \\
$[NeII]$	& 12.80	& 0.22	& \\
N8.6	& 8.6	& 0.43	&  PAH1 \\
N8.7	& 8.78	& 1.75	& Short N-band \\
SiC	& 11.81	& 2.36 & \\	
Nband	& 10.40	& 5.25	& Wide N-band \\
\end{tabular}
\end{center}
\caption{\label{tab_MIDIfilters}Different spectral filters, which can be used in the MIDI target acquisition mode to record images with one UT.}
\end{table}

Of course there is no general answer to this topic. But if the correlated flux density of the science target is in general high enough for MIDI, often also without tip/tilt-correction a fringe measurement has been proven to be feasible.
Further, if no fringes are detectable due to low intrinsic source visibility, the high-quality state-of-the-art equipment of MIDI offers currently the possibility, to record target acquisition images with the MIDI science detector applying wide- and different narrow-band filters (see Table~\ref{tab_MIDIfilters}). The FWHM of the single-telescope PSF can be expected to be about 250~mas.

The concrete strategies to fully exploit the actual MIDI capabilities as discussed above
will be fixed and the outcome will lead to 'Notes for MIDI astronomers' which should help in the 
preparation and conduction of non-standard observations, due to the faint and/or complex source 
structure. A first version is already available at the Paranal Observatory. For more information or 
a copy please contact J.-U.~Pott.

\subsection{AMBER}
The near infrared focal instrument AMBER of the VLTI combines three beams in J, H and K band (1-2.5~$\mu$m). Without external fringe tracker the limiting  K-band magnitudes are K$\sim$13 for the use of the 8~m Unit Telescopes (UT) and K$\sim$10 for the use of the 1.8~m Auxiliary Telescopes (AT). If a bright (H$<$13) reference star is present within 1~arcmin distance to the science target, the limiting magnitudes may rise up to K$\sim$20 (UTs) and K$\sim$16 (ATs), depending on the observing mode. The maximum spectroscopic resolution is foreseen to reach 10\,000. While AMBER is currently under commissioning, the public scientific usage is foreseen to start in April 2005. Details on the current instrument status are given in \cite{Glindemann04} and references therein. 

In the next section we present VLTI observations of the Galactic Center region in the near infrared. They will test the capabilities of AMBER in the densely populated GC region in a similar way as pointed out in detail Sect.~\ref{sect_MIDIcaps} for the MIDI observations. 

\subsubsection{Stellar Orbits}
The actual measurements of mass and density of the supermassive black hole in the Galactic 
Center are based on dynamic measurements of the stars orbiting the black hole (e.g.\
 \cite{2002Natur.419..694S,2003ApJ...596.1015S}). The high angular resolution data was obtained using Speckle techniques (res$\sim$100~mas 
in K, Sharp at the NTT) and later on Adaptive Optics (res$\sim$60~mas in K, NACO at the VLT-UT4). The 
precision of the orbital parameters, derived from the maps and spectra, can be significantly 
improved by K-band interferometric images, obtained with the 3-beam combining device AMBER. Using 
closure phases the original brightness distribution can be restored. Within the AMBER field-of-view of $\sim$0.25`` at 2$\mu$m, employing the ATs (the FOV is given by the size of the Airy Disks of the single telescopes), a number of sources can be observed in the Galactic Center region. Even the infrared flares of the accretion disk around the black hole at maximum emission were shown to be as bright as some of the stars\cite{2003Natur.425..934G} and therefore well above the magnitude limit for AMBER observations with an available bright off-axis reference source.

With the maximum AT separation 
of about 200~m the interferometric angular resolution in K-band will be about 2~mas. Thus the 
resolution may rise about one order of magnitude employing the VLTI. With higher precision in the stellar motion data 
the ambiguity between Keplerian and non-Keplerian orbit fits to the data may be resolvable (cf. 
\cite{Mouawad2004}). Thus the investigation of the central gravitational potential will benefit a 
lot from the application of VLTI-AMBER. 

\subsection{\label{sect_massivestars}The cluster of massive stars}

The luminosity of the entire central parsec of our Galaxy is dominated by a cluster of 
massive stars, formed only a few million years ago (e.g.\ \cite{1997A&A...325..700N}). They observed strong 
stellar winds ($\dot{M}\sim (5-80)\cdot 10^{-5} M_\odot/yr$) with relatively small outflow 
velocities (V$\sim$300 to 1000~km/s). These findings together with effective temperatures 
(17-30$\cdot 10^3$K) and strong enhanced helium abundances ($N_{He}/N_H>0.5$) let them conclude, 
that the observed HeI emission line stars are evolved  blue supergiants. These stars appear to be 
close to the Ofp9/Wolf-Rayet evolutionary stage and power the central cluster. Because of the huge source 
density in the central stellar cluster, single-telescope resolutions of a few hundred mas cannot 
distinguish between close double/multiple stars systems. The linear scale of 
(0.5"$\sim20~mpc\sim4\cdot10^3~AU$ at a Distance of $\sim 8~kpc$) is huge with respect to the 
stellar radii of HeI~stars. The K-band resolution of AMBER is needed to resolve these sources, if 
they are in binary systems (e.g.\ IRS~16SW in \cite{1999ApJ...523..248O}). The correct number of stars is needed to derive 
the number of ionizing photons and thus understand, if the massive stars can account for the entire 
observed HeI continuum radiation of the Galactic~Center. Binarity and mass transfer between components would also help to explain the unexpectedly large number of He-stars at the center. 

\section{Summary}

It was shown, that in the Galactic Center region numerous scientifically interesting targets exist. Their observation will at the same time reveal unprecedented scientific results as well as test the capabilities of VLTI on faint, embedded targets, partially providing properties, which are not ideal for interferometric observations. Thus valuable experience for future VLTI observations will be made. 

We presented the preparation of VLTI/MIDI observations of the enigmatic dust embedded sources in the Galactic Center in detail and discussed thereby different stumbling blocks, which may hamper or even inhibit successful observations of non-standard targets. The results were rather encouraging. It was shown, that blind target acquisition is efficiently feasible, if coordinates of intermediate precision (subarcsec resolution is provided by all modern MIR observation with large single-telescope observations) are given. Several other aspect were mentioned to give the reader an idea of the feasibility of VLTI observations, close to the offered capabilities.

The experiences with MIDI observing faint targets will help to conduct observations on extragalactic targets in the near future, because the properties of extragalactic targets are as well close to MIDI's (current) instrumental capabilities.

Because of the permanent expansion of the VLTI in general and of the ongoing commissioning runs of the science instruments and other VLTI facilities, the capabilities of the VLTI, offered to the astronomical community, will be enhanced continuously. Therefore interested astronomers are requested to contact the authors and respective instrument responsibles for most recent information.

\acknowledgments     
We thank Anders~Wallander and Sebastien Morel for substantial discussions. This work has been partly supported by the
Deutsche Forschungsgemeinschaft (DFG)
via SFB494 at the University of Cologne.

\bibliography{SPIEfease}
\bibliographystyle{spiebib}

\end{document}